\documentclass[aip,apl,amsmath,amssymb,
reprint,%
nofootinbib,
]{revtex4-2}
\usepackage{lmodern}
\usepackage[bookmarks,%
            colorlinks,%
            urlcolor=blue,%
            citecolor=blue,%
            linkcolor=blue,%
            hyperfigures,%
            pdfcreator=KN,%
            breaklinks=false]{hyperref}
\usepackage[english]{babel}
\usepackage{siunitx}
\usepackage{csquotes}
\usepackage{nth}
\usepackage{multirow}
\usepackage[version=4]{mhchem}
\usepackage[caption=false]{subfig}
\usepackage{graphicx}
\usepackage{orcidlink}
\DeclareGraphicsExtensions{.pdf}
\begin{document}

\title{Effect of thermal annealing on dielectric and ferroelectric properties of aerosol-deposited \ce{0.65Pb(Mg_{1/3}Nb_{2/3})O_{3}-0.35PbTiO3} thick films}
\author{Kevin Nadaud\orcidlink{0000-0002-2969-1453} \textsuperscript{a,b)}}
\affiliation{%
GREMAN UMR 7347, Université de Tours, CNRS, INSA-CVL, 16 rue Pierre et Marie Curie, 37071 Tours, France%
}%

\author{Matej Sadl \orcidlink{0000-0003-1497-5610}\textsuperscript{a)}}
\affiliation{%
Electronic Ceramics Department, Jožef Stefan Institute, Jamova cesta 39, 1000 Ljubljana, Slovenia%
}
\affiliation{%
Jožef Stefan International Postgraduate School, Jamova cesta 39, 1000 Ljubljana, Slovenia%
}
\author{Micka Bah\orcidlink{0000-0001-6636-2854}}
\affiliation{%
GREMAN UMR 7347, Université de Tours, CNRS, INSA-CVL, 16 rue Pierre et Marie Curie, 37071 Tours, France%
}

\author{Franck Levassort\orcidlink{00000-0002-2019-9390}}
\affiliation{%
GREMAN UMR 7347, Université de Tours, CNRS, INSA-CVL, 16 rue Pierre et Marie Curie, 37071 Tours, France%
}

\author{Hana Ursic\orcidlink{0000-0003-4525-404X}}
\affiliation{%
Electronic Ceramics Department, Jožef Stefan Institute, Jamova cesta 39, 1000 Ljubljana, Slovenia%
}
\affiliation{%
Jožef Stefan International Postgraduate School, Jamova cesta 39, 1000 Ljubljana, Slovenia%
}

\keywords{Impedance spectroscopy, hyperbolic analysis, aerosol deposition, PUND measurements, domain wall}

\begin{abstract}
In this work, the effects of thermal annealing at \qty{500}{\celsius} on aerosol-deposited \ce{0.65Pb(Mg_{1/3}Nb_{2/3})O_{3}-0.35PbTiO3} thick films on stainless-steel substrates are investigated using two complementary methods at high and low applied external electric fields.
The first one is Positive Up Negative Down method, which allows us to obtain information about the switching and non-switching contributions to the polarization.
It shows that the as-deposited film is ferroelectric before annealing, since it has a switching contribution to the polarization.
After annealing, both the switching and non-switching contributions to polarization increased by a factor of 1.6 and 2.33, respectively, indicating stronger ferroelectric behavior.
The second method is based on impedance spectroscopy coupled with Rayleigh analysis.
The results show that post-deposition thermal annealing increases the reversible domain wall contribution to the dielectric permittivity by a factor 11 while keeping the threshold field similar.
This indicates, after annealing, domain wall density is larger while domain wall mobility remains similar.
These two complementary characterization methods show that annealing increases the ferroelectric behavior of the thick film by increasing the domain wall density and its influence is visible both on polarization versus electric field loop and dielectric permittivity.

\end{abstract}
\def\thefootnote{a)}\footnotetext{These authors contributed equally to this work}
\def\thefootnote{b)}\footnotetext{Author to whom correspondence should be addressed: \href{mailto:kevin.nadaud@univ-tours.fr}{kevin.nadaud@univ-tours.fr}}
\def\thefootnote{c)}\footnotetext{The following article has been accepted by Applied Physics Letters. It can be found at \href{https://doi.org/10.1063/5.0087389}{https://doi.org/10.1063/5.0087389}}
\def\thefootnote{\arabic{footnote}}
\maketitle


Thanks to their large relative dielectric permittivity and polarization, relaxor-ferroelectrics are promising materials for energy storage devices.
Even though ferroelectric capacitors have lower energy density than batteries and supercapacitors, they can charge/discharge under large currents and can be integrated into compact pulsed-power and power-conditioning electronic devices using thin or thick films \cite{PalneediAFM2018}.
One of the most promising relaxor-ferroelectrics is \ce{0.65Pb(Mg_{1/3}Nb_{2/3})O3-0.35PbTiO3} (PMN--35PT) ceramics with a morphotropic phase boundary composition that exhibits excellent dielectric, piezoelectric and ferroelectric properties.\cite{UrsicJJAP2011,KellyJACS2005,AlgueroJACS2005}

A recent aerosol deposition method has opened new opportunities in device engineering of rapidly advancing thick film technologies. 
The advantage of aerosol deposition is the deposition of highly dense and crack-free thick films at room temperature.\cite{AkedoJTST2008,HanftJCST2015}
Since no high temperature processing is required, material compatibility is superior and allows integration of ceramics with new substrates that are morphologically unstable at high temperatures, such as polymers and metals.\cite{Kosec2011}
In the aerosol deposition method, the powder containing micrometre-sized particles is accelerated to velocities between \qty{150}{\m\per\s} and \qty{500}{\m\per\s}, under vacuum conditions before hitting the substrate \cite{AkedoJACS2006}. 
Therefore, film deposition and  growth occur as a result of sufficiently high kinetic energy of the impacting particles, which is converted into fracture energy, consolidation and plastic deformation \cite{ExnerJECS2019}.  
The aerosol deposition produces ceramic thick films with properties inherently different from conventional ceramics prepared by sintering. After aerosol deposition, the thick films have high density (over \qty{95}{\percent} of the theoretical density) \cite{AkedoMSF2004}, nano-sized pores \cite{FanJTST2007,SadlAM2021}, and good adhesion to the substrate. 
The typical microstructure of aerosol-deposited ceramic films shows grains reaching only a few tens of nanometers in diameter \cite{AkedoJTST2008,HenonJECS2015}. 
The grain size of the ceramics is known to strongly influence the dielectric and piezoelectric properties of ferroelectric material\cite{chengapl1999,BassiriGharbJE2007,SchultheiJECS2020}.
Moreover, Damjanovic \emph{et al} also show that the irreversible contribution of domain walls is smaller for small grains than for coarse grains \cite{DamjanovicJPCM1997}.
In addition, the impacts generate internal compressive stresses that can be on the order of several hundred MPa to several GPa \cite{KhansurSM2018,SchubertM2014}. 
Both characteristics, reduced grain/crystallite size and internal stresses can be especially detrimental for as-deposited ferroelectric thick films and their functional properties (e.g. dielectric, piezoelectric and ferroelectric properties). To improve the functional properties, ferroelectric thick films are often subjected to a moderate thermal annealing. 
Already at temperatures of \qty{500}{\celsius}, a significant stress-relaxation occurs, which is believed to be responsible for the enhancement of the ferroelectric response \cite{KhansurSM2018}.

Electrical characterization methods of ferroelectric materials can generally be divided into high (supercoercive) field and low (subcoercive) field measurements.
High field range is considered when the applied electric field is three times the coercive field, and a low field range when the applied electric field is less than the half the coercive field\cite{GharbJAP2005}.
The Positive Up Negative Down method (PUND) \cite{WangMD2020,LiuAPL2016,MartAPL2019} measurement technique complements conventional polarization versus electric field ($P(E)$) loop measurements by deconvolving the switching.
It consists of applying successive voltage pulses (pre-polarization then P, U, N and D for positive, up, negative, and down, respectively) to a ferroelectric capacitor, and recording the current flow during these pulses.
Positive and up voltage pulses have the same polarity, positive with respect to the bottom electrode which is grounded, whereas negative and down voltage pulses have both negative polarity.
For {P} and {N} pulses, the current is the sum of the different contributions (i) leakage, (ii) capacitive and (iii) switching, since the previous pulse had the opposite polarity.
For {U} and {D} pulses, the current is the sum of only two contributions (i) leakage and (ii) capacitive, since the previous pulse had the same polarity.
By subtracting the currents $i_{P} - i_{U}$ and $i_{N} - i_{D}$, it is possible to extract only the switching contributions for both polarities.
Then numerical integration over time is used to obtain the $P(E)$ loop for the switching contribution only.

Most often, low-field measurements are based on impedance spectroscopy, which can be performed, for example, as a function of frequency and temperature to reveal relaxor phenomena\cite{GonzlezAbreuJAC2019,MirandaJECS2001} or DC bias field to determine the tunability of the material\cite{NadaudJAP2016,GartenJAP2014}.
In ferroelectric materials, the irreversible domain wall process contributes significantly to the relative permittivity at sub-coercive driving fields\cite{taylorjap1997,borderonapl2011}.
Thus, measurement as a function of driving field enables characterization of domain wall motions that depend on the structure of the material\cite{NadaudJAP2015,NadaudAPL2021,GharbJAP2005,GartenJAP2014}.

In this paper, the effect of the post-deposition by thermal annealing aerosol-deposited PMN--35PT thick films is investigated using methods at high and low fields.
PUND analysis is used to characterize the magnitude of the ferroelectric switching component to the measured polarization before and after thermal treatment.
The low field method is based on impedance spectroscopy as a function of driving field and as a function of frequency to find a dielectric relaxation.
This is coupled with the Rayleigh analysis to further investigate domain wall mobility in the as-deposited and annealed films.

\begin{figure}
    {\includegraphics[width=0.45\textwidth]{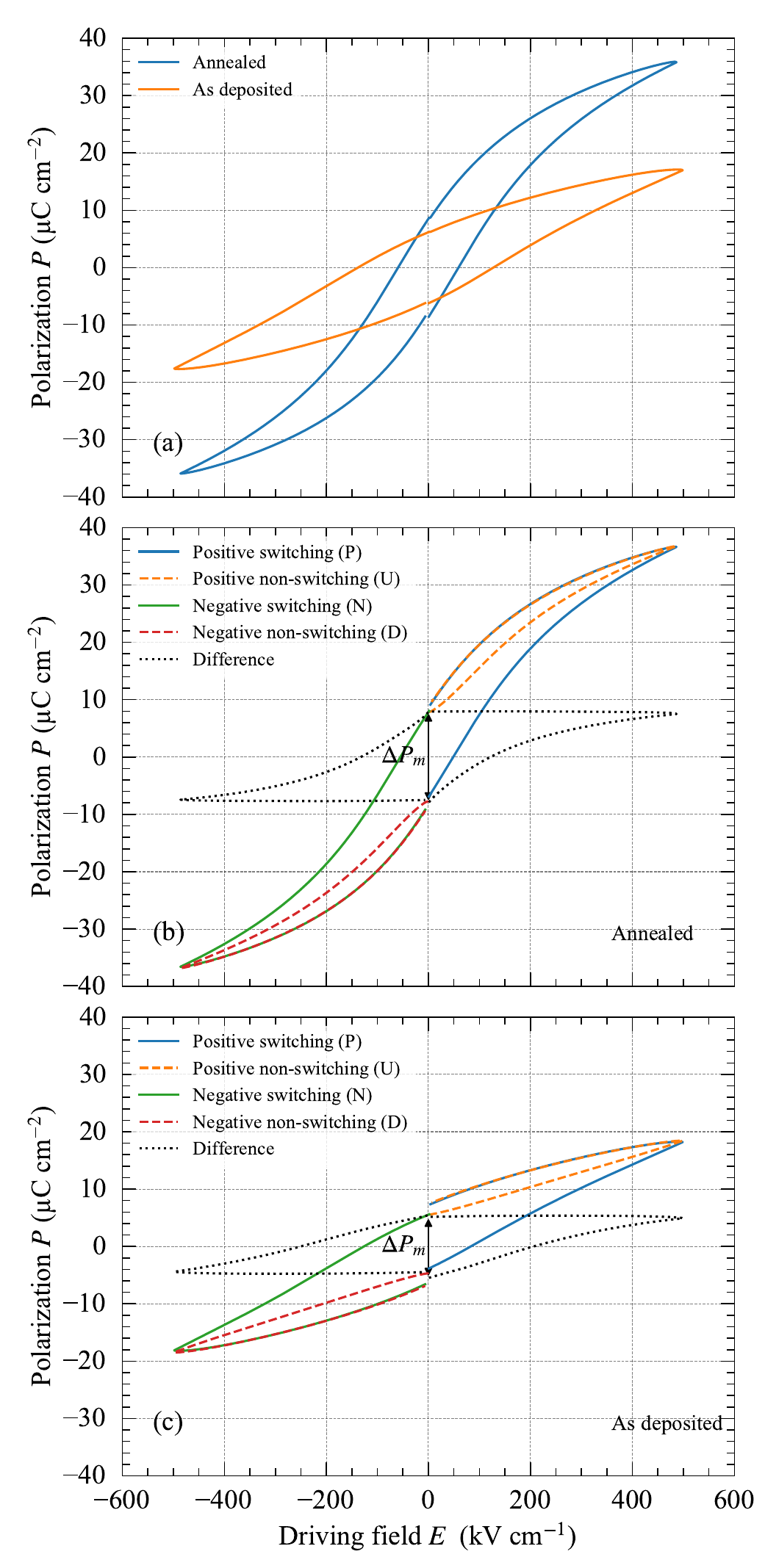}%
    \subfloat{\label{subfig:PE AsDep Ann}}%
    \subfloat{\label{subfig:PE Ann PUND}}%
    \subfloat{\label{subfig:PE AsDep PUND}}}
    \caption{Polarization versus electric field loops for annealed and as-deposited samples (a). PUND measurement for the annealed sample (b) and for the as-deposited sample (c). The black dotted curve in (b) and (c) indicates the switching contribution to the polarization.}
    \label{fig:PE}
\end{figure}

\ce{PbO} (\qty{99.9}{\percent}, Aldrich), \ce{MgO} (\qty{99.95}{\percent}, Alfa Aesar), \ce{TiO2} (\qty{99.8}{\percent}, Alfa Aesar) and \ce{Nb2O5} (\qty{99.9}{\percent}, Aldrich) were used for the synthesis of \ce{0.65Pb(Mg_{1/3}Nb_{2/3})O_{3}-0.35PbTiO3} powder.
First, two powder mixtures corresponding to PMN and PT were homogenized separately. 
Then, the powder mixture corresponding to PT was calcined at \qty{750}{\celsius} for \qty{2}{h} to facilitate faster reaction kinetics.\cite{DragomirJECS2019} 
Both powder mixtures were homogenized together and reacted for \qty{24}{h} in the mechanochemical-activation-assisted synthesis. 
The powder was then milled for \qty{2}{h}, annealed at \qty{900}{\celsius} for \qty{1}{h} and finally milled for \qty{0.5}{h}. 
The powder was deposited onto \qtyproduct{15x15x0.8}{\mm} stainless-steel substrates (no. 304, American Iron and Steel Institute) at room temperature using aerosol deposition method.
Full details of powder processing and aerosol deposition conditions are described elsewhere \cite{SadlAM2021}.
A schematic of the aerosol deposition apparatus is represented in ref. \onlinecite{SadlJMECM2019}.
The as-deposited thick film samples were annealed at \qty{500}{\celsius} for \qty{1}{h} using \qty{2}{\K\per\minute} heating and cooling rates under an air atmosphere.
The as-deposited and annealed films are two distinct samples with film thickness of \qty{3.6}{\micro\m} and \qty{4.1}{\micro\m}, respectively.
For electrical characterization, gold was sputtered through a shadow mask to form circular top electrodes with diameter of \qty{0.5}{\mm}.
The stainless-steel substrate acted as a bottom electrode.

Polarization vs. electric field ($P(E)$) and PUND measurements were performed using the AixACCT TF2000 ferroelectric analyzer.
The current is measured as a function of time while a triangular waveform is applied with a slew rate of \qty{4}{\kV\per\s} and the delay between pulses is \qty{1}{\micro\s}.
The polarization is computed by the TF2000 analyzer software by numerical integration. 

The impedance measurement was performed using an Agilent 4294A impedance analyzer.
The relative dielectric permittivity $\varepsilon_{r}'$ of the material inside a metal-insulator-metal topology was calculated using the measured capacitance $C$, based on the parallel plates formula: 
\begin{equation}
    \varepsilon_{r}' = \frac{t}{S\varepsilon_{0}}C.
\end{equation}
with $\varepsilon_{0} =$ \qty{8.85e-12}{\farad\per\m} the dielectric vacuum permittivity, $S$ the area of the top electrode (\qty{0.284}{\mm\squared}) and $t$ the thickness of the material (\qty{3.6}{\micro\m} for the as-deposited sample and \qty{4.1}{\micro\m} for the annealed sample, respectively).
The imaginary part of the permittivity is given by:
\begin{equation}
    \varepsilon_{r}'' = \varepsilon_{r}'\tan\delta
\end{equation}
With $\tan\delta$ the dielectric losses.
All electrical characterizations were performed at room temperature, i.e. $T=\qty{25}{\celsius}$.
For the impedance measurement as function of frequency, the driving field is $\qty{3.4}{\kV\per\cm}$ and for the Rayleigh analysis, the driving field goes from \qty{0.034}{\kV\per\cm} to \qty{3.4}{\kV\per\cm}. 

The structural and microstructural analyses of the as-deposited and annealed PMN--35PT thick films were done with X-ray diffraction (XRD) and scanning electron microscopy (SEM) and are presented in the supplementary material. 
In summary, the results reveal no microstructural changes after annealing of the thick films. 
According to SEM the density and porosity remain the same. 
In addition, XRD analysis shows that annealing has a minor influence on crystallite size, but significantly (\qty{41}{\percent}) reduces the microstrain. 
Similar observations were made for other aerosol-deposited thick films. 
Annealing at moderate temperatures usually reduces the internal stresses, but the microstructure is preserved \cite{SadlAM2021,KhansurJCI2018}

Fig.~\ref{subfig:PE AsDep Ann} shows the $P(E)$ loops for the annealed and the as-deposited samples. 
The maximum polarization is larger for the annealed sample (\qty{36}{\micro\coulomb\per\centi\metre\squared} vs. \qty{17}{\micro\coulomb\per\centi\metre\squared}).
The polarization value at maximum electric field (\qty{480}{\kV\per\cm}) for as-deposited and annealed samples is similar to that reported by Park \emph{et al} \cite{ParkCI2018}.
To determine the origin of this larger maximum polarization, a PUND measurement was performed and the results are shown in Figs.\ref{subfig:PE Ann PUND} and \ref{subfig:PE AsDep PUND}.
For both samples, the difference in polarization between switching and non-switching polarization yields the switching contribution, which is represented by the  PUND-corrected loops (dotted black curves). 
Such loops have a typical shape, they have straight horizontal lines when the field returns to zero\cite{WangMD2020}.
The difference between maximum and minimum polarization, $\Delta P_{m}$, was calculated and the values are \qty{15.6}{\micro\coulomb\per\centi\metre\squared} and \qty{9.8}{\micro\coulomb\per\centi\metre\squared} for the annealed and as-deposited samples, respectively.
The higher value for the annealed sample is consistent with the larger value of polarization related to stress relaxation of compressive in-plane stress \cite{SadlAM2021}.
Nevertheless, the as-deposited sample is also ferroelectric, even though higher stresses in the film decrease the maximum polarization and increase the coercive field. 
Similar observation of the decrease of coercive field after thermal annealing were reported also in other aerosol deposited PMN--35PT thick films on Si substrates, while the actual origin was not precisely determined\cite{ParkCI2018}.

\begin{figure}[t]
    \includegraphics[width=0.45\textwidth]{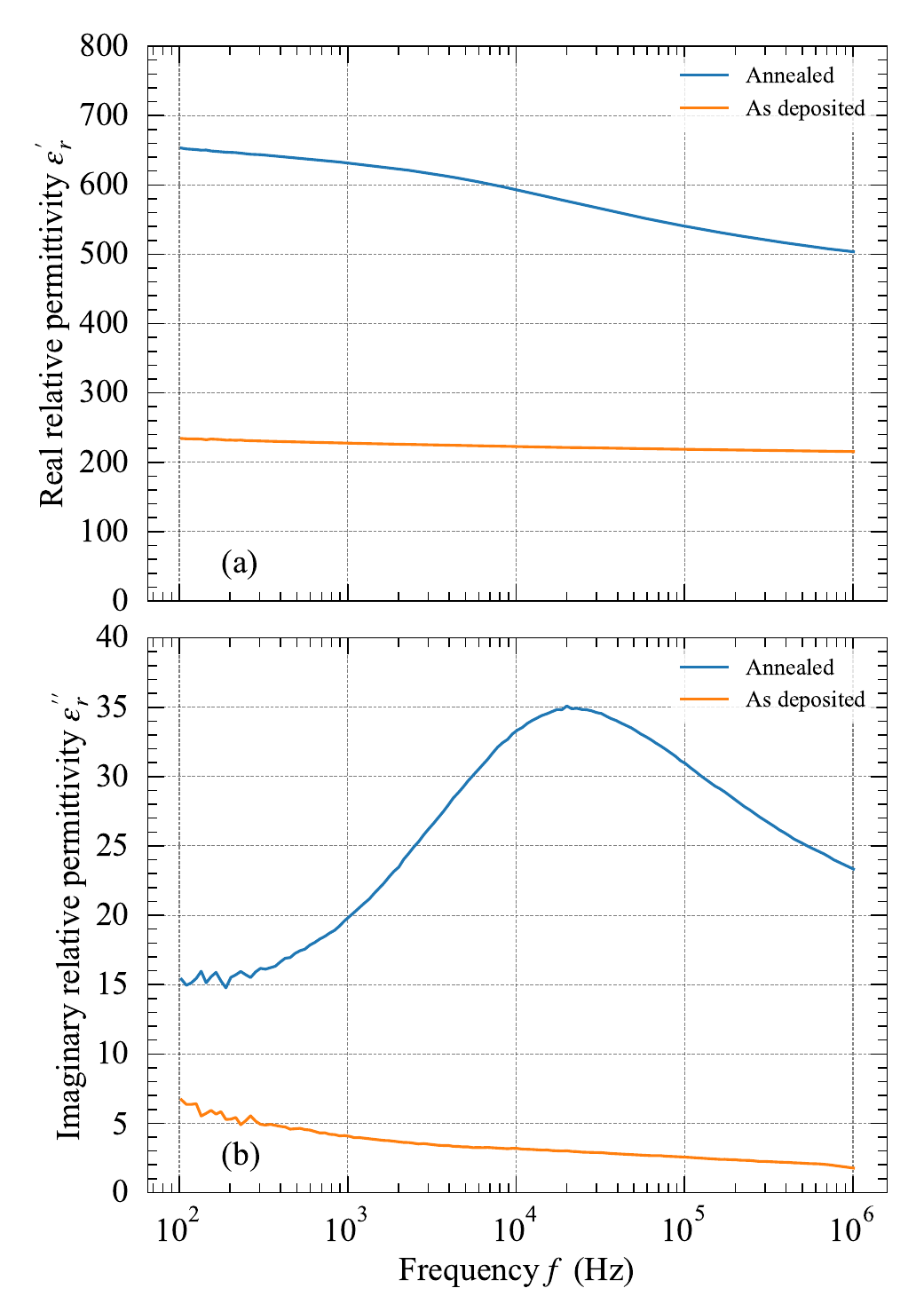}
    \caption{Real (a) and imaginary (b) parts of the relative permittivity, for the as-deposited and the annealed samples, as a function of the frequency.}
    \label{fig:realPerm imagPerm AsDep Ann FREQ}
\end{figure}

Fig.~\ref{fig:realPerm imagPerm AsDep Ann FREQ} shows the real and imaginary parts of the relative permittivity as a function of frequency for a driving field of $\qty{3.4}{\kV\per\cm}$.
The real part of the permittivity of the annealed sample is about three times higher than for the as-deposited sample (540 vs. 219 at \qty{100}{\kHz}), showing the strong effect of annealing on the dielectric properties.
The higher value of relative permittivity when the stress is reduced is similar to what has been reported for PMN-30PT \cite{KeechJAP2014} or for PZT \cite{GriggioPRL2012} thin films.
This large difference between annealed and as-deposited films is also visible in the imaginary part of the relative permittivity, 31.9 vs. 2.57 at \qty{100}{\kHz}, corresponding to a dielectric losses of 0.057 and 0.012, respectively.
The much larger permittivity and dielectric losses are consistent with a higher ferroelectric contribution in the annealed sample found in the PUND measurements in the previous part.

\begin{figure}[t]
    {\includegraphics[width=0.45\textwidth]{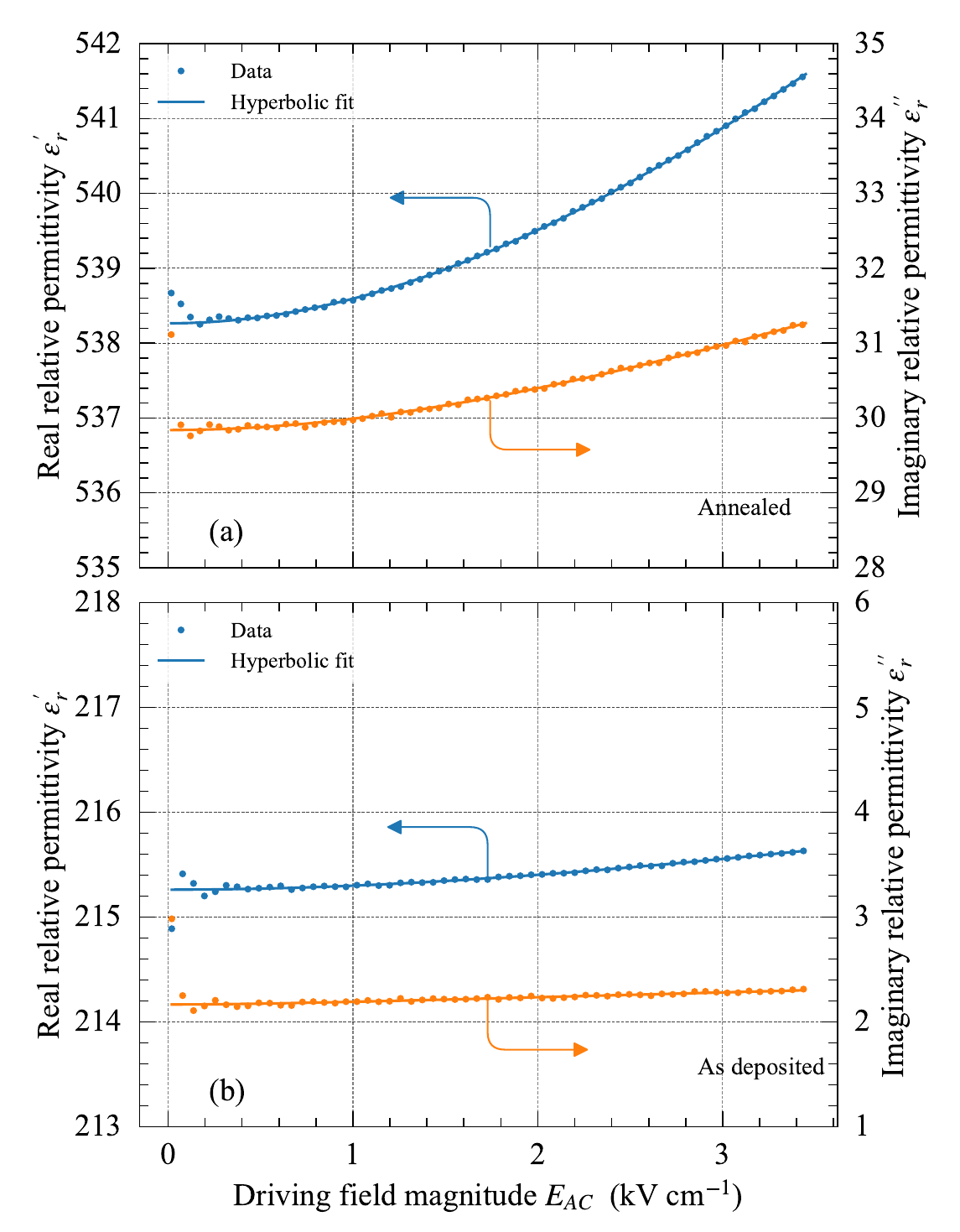}%
    \subfloat{\label{subfig:realPerm imagPerm Ann OLEV}}%
    \subfloat{\label{subfig:realPerm imagPerm AsDep OLEV}}}
    \caption{Real and imaginary parts of the relative permittivity as a function of the driving field magnitude, for the annealed sample (a) and for the as-deposited sample (b).}
    \label{fig:realPerm imagPerm As Dep Ann OLEV}
\end{figure}

Another notable difference in relative permittivity concerns its frequency dependence. 
The imaginary part of the relative permittivity of the annealed sample exhibits a clearly visible maximum at about \qty{20}{\kHz}, corresponding to dielectric relaxation\cite{GonzlezAbreuJAC2019,MirandaJECS2001}, while for the as-deposited sample the imaginary part decreases with frequency and has no local maximum.
For the real part, the decrease with frequency is more pronounced for the annealed sample, 652 to 502 from \qty{100}{\Hz} to \qty{1}{\MHz} than for the as-deposited sample, 233 to 216, which is also due to dielectric relaxation around \qty{20}{\kHz}.

The measurement of relative permittivity as a function of the driving electric field, which allows to deconvolve the different contributions to permittivity\cite{damjanovicrpp1998,borderonapl2011,NadaudAPL2021} (lattice, reversible and irreversible domain wall contributions), was performed for a fixed frequency of \qty{100}{\kHz} and the results are shown in Fig.~\ref{fig:realPerm imagPerm As Dep Ann OLEV}.
The real and imaginary parts of the relative permittivity follow the generalized Rayleigh law, named hyperbolic law:\cite{borderonapl2011,borderonapl2014,NadaudAPL2021,NadaudJAP2016}
\begin{equation}\label{eq:hyperbolic law}
    \varepsilon_{r} = \varepsilon_{\mathit{r-l}} + \sqrt{\varepsilon_{\mathit{r-rev}}^{2}+(E_{\mathit{AC}}\alpha_{r})^{2}}.
\end{equation}
Where $\varepsilon_{\mathit{r-l}}$ is the lattice contribution to the permittivity, $\varepsilon_{\mathit{r-rev}}$ and $\alpha_{r}$ correspond to the reversible domain wall contribution, also called vibrations, and irreversible domain wall contribution, also called domain wall pinning/unpinning, respectively.
$E_{\mathit{AC}}$ corresponds to the magnitude of the applied driving electric field.
The Rayleigh parameter $\alpha_{r}$ corresponds to the slope of the asymptote at a high electric field.
The higher the value, the higher is the irreversible domain wall contribution.
When the electric field increases from \qty{0.034}{\kV\per\cm} to \qty{3.4}{\kV\per\cm}, in the annealed sample, the real part increases from 538.2 to 541.6 and the imaginary part from 29.8 to 31.3, which is due to the irreversible domain wall contribution.
This increase is almost not visible in the as-deposited sample indicating a very small irreversible domain wall contribution since the increases for real and imaginary parts are less than 0.2.
The real and imaginary parts of the permittivity were fitted (using the Levenberg-Marquat method\cite{lmfit}) to equation \eqref{eq:hyperbolic law} to extract the coefficients, which are listed Table~\ref{tab:hyperbolic coefficients}.

The real and imaginary parts of the lattice contribution $\varepsilon_{\mathit{r-l}}'$, $\varepsilon_{\mathit{r-l}}''$, which is the main contribution to the permittivity follow the same trend as the total permittivity given Fig.~\ref{fig:realPerm imagPerm AsDep Ann FREQ}, i.e. larger values for the annealed sample.
The irreversible domain wall contribution, represented by the Rayleigh parameter $\alpha_{r}$, is strongly affected by annealing.
In the annealed film, the value of the real part is 10 times higher (\qty{2.38(3)}{\cm\per\kV} vs. \qty{0.245(9)}{\cm\per\kV}) and the value of the imaginary part is 18 times higher (\qty{0.86(2)}{\cm\per\kV} vs. \qty{0.048(2)}{\cm\per\kV}) than the values of the as-deposited thick film.
This large difference corresponds to the slope difference observed in Fig.~\ref{fig:realPerm imagPerm As Dep Ann OLEV}.
The higher value of the irreversible domain wall contribution when the stress is reduced is similar to what has been reported for PMN--30PT\cite{KeechJAP2014} or for PZT \cite{GriggioPRL2012} thin films.

The reversible domain wall contribution, which is proportional to the domain wall density\cite{boserjap1987,BorderonSR2017,NadaudJAP2016}, is also affected by annealing.
For the annealed sample, the real part of the reversible domain wall contribution $\varepsilon_{\mathit{r-rev}}'$ is 11 times higher than for the as-deposited sample indicating a much larger domain wall density for the annealed sample.
Those findings are supported also by the PFM analysis (Supplementary material S3). 
There are no studies that would directly correlate between the stress relaxation and domain wall density. 
However, the phase composition of the PMN--35PT solid solution at the morphotropic phase boundary (MPB) is very sensitive to the presence of the stress. 
This has been demonstrated in screen-printed PMN--35PT thick films\cite{UrsicJECS2010,UrsicJAP2011}. 
By using Rietveld refinement analysis, it was shown that the ratio between the monoclinic ($Pm$) and tetragonal ($P4mm$) phase varies with the magnitude of in-plane stress in PMN–35PT thick films. 
Therefore, the change in the stress magnitude promotes polarization rotation and thus phase transformation in PMN-35PT thick films, which was also previously observed for bulk PMN--PT \cite{HouPRB2018} and other bulk materials at MPB\cite{DamjanovicAPL2010}.
In contrast, the aerosol-deposited thick films exhibit XRD peak broadening typical for this deposition method, which complicates the determination of the phase composition of the crystal phases. 
Nevertheless, we can assume that the stress relaxation in aerosol-deposited films also induces phase transformation in the MPB compositions, which leads to a change in domain structure affecting also the domain wall density. 


\begin{table}
    \caption{Numerical values of the extracted coefficients using the hyperbolic laws of the curves shown in Fig.~\ref{fig:realPerm imagPerm As Dep Ann OLEV}. The associated uncertainties are determined during the fitting process and correspond to the confidence intervals.}
    \label{tab:hyperbolic coefficients}
    \resizebox{0.45\textwidth}{!}{%
    \begin{tabular}{llS[table-figures-decimal = 4]S[table-figures-decimal = 4]}
    \hline
    \hline
    &              & \multicolumn{2}{c}{Conditions}\tabularnewline
    \cline{3-4}
    & Contribution & {Annealed }     & {As deposited} \tabularnewline
    \hline
    \multirow{3}{*}{Real part}           & $\varepsilon_{r-l}'$             & 529.8(3)  & 214.49(7) \tabularnewline
                                         & $\varepsilon_{r-rev}'$           & 8.5(3)    & 0.78(7)   \tabularnewline
                                         & $\alpha_{r}'$ (\unit{\cm\per\kV})  & 2.38(3)   & 0.245(9)  \tabularnewline
    \hline
    \multirow{3}{*}{Imaginary part}      & $\varepsilon_{r-l}''$            & 27.5(2)   & 2.135(7)  \tabularnewline
                                         & $\varepsilon_{r-rev}''$          & 2.4(2)    & 0.03(1)   \tabularnewline
                                         & $\alpha_{r}''$ (\unit{\cm\per\kV}) & 0.86(2)   & 0.048(2)  \tabularnewline
    \hline
    \multirow{3}{*}{Dissipation factors} & $m_{\varepsilon_{r-l}}$          & 0.0519(3) & 0.0099(1) \tabularnewline
                                         & $m_{\varepsilon_{r-rev}}$        & 0.28(2)   & 0.04(1)   \tabularnewline
                                         & $m_{\alpha_{r}}$                 & 0.36(1)   & 0.20(1)   \tabularnewline
    \hline
    Threshold field                      & $E_{th}$ (\unit{\kV\per\cm})       & 3.56(7)   & 3.2(2)    \tabularnewline
    \hline
    \hline
    \end{tabular}}
\end{table}

Based on the reversible and irreversible domain wall contributions, it is possible to calculate the threshold field:\cite{BorderonSR2017,borderonapl2011}
\begin{equation}
    E_{\mathit{th}} = \frac{\varepsilon_{\mathit{r-rev}}'}{\alpha_{r}'}
\end{equation}
The threshold field represents the degree of domain wall pinning in the material\cite{GharbJAP2005}.
For the two samples, the values are very similar, \qty{3.56(7)}{\kV\per\cm} for the annealed sample and \qty{3.2(2)}{\kV\per\cm} for the as-deposited sample.
This means that annealing does not change the depth of the pinning center in the material and the difference in terms of the driving field sensitivity, represented by the Rayleigh parameter $\alpha_{r}$, is due to the different number of domain walls.
The values obtained are of the same order of magnitude as in \ce{0.5Pb(Yb_{1/2}Nb_{1/2})O3 -0.5PbTiO3} \cite{GharbJAP2005} ($E_{\mathit{th}} = \qty{2.2}{\kV\per\cm}$), \ce{Pb(Zr_{0.57}Ti_{0.43})O3}\cite{borderonapl2014} ($E_{\mathit{th}} = \qty{2.7}{\kV\per\cm}$) or well-oriented \ce{Ba_{2/3}Sr_{1/3}TiO3}\cite{NadaudAPL2018} ($E_{\mathit{th}} = \qty{1.9}{\kV\per\cm}$), indicating low depth of the pinning centers and a high mobility of the domain walls.

Using the real and imaginary parts of the individual contribution, the associated dissipation factor can be calculated using:
\begin{equation}
    m_{x} = \frac{x''}{x'}
\end{equation}
with $x'$ and $x''$ are the real and imaginary parts respectively of $\varepsilon_{\mathit{r-l}}$, $\varepsilon_{\mathit{r-rev}}$ and $\alpha_{r}$. 
For the lattice contribution, the $m_{\mathit{r-l}}$ value for the annealed sample is 5 times higher, which is in agreement with the observations on the whole permittivity and is due to the dielectric relaxation around \qty{20}{\kHz}.
It can seen that the dissipation factor for the irreversible domain wall contribution $m_{\alpha_{r}}$ is one order of magnitude higher than for the lattice contribution.
After annealing, the dissipation factor is higher due to the higher density of the domain walls, as there are more interactions between the domain walls\cite{NadaudJAP2016,BorderonSR2017}.
For the reversible domain wall contribution, the dissipation factor is 8 times higher for the annealed sample than for the as-deposited sample. 
This large difference is again attributed to the interaction between the domain walls which greatly affects the dissipative behavior of this contribution\cite{NadaudJAP2016,BorderonSR2017}, since before annealing the low domain wall density results in a small interaction and after annealing the large domain wall density results in much larger interactions.


To summarize, in this study, the effect of thermal annealing at \qty{500}{\celsius} of PMN--35PT thick films was characterized using methods at high field ($P(E)$ and PUND) and low field (impedance spectroscopy).
The measurement of $P(E)$ loops shows that annealing increases the maximum polarization from \qty{17}{\micro\coulomb\per\centi\metre\squared} to \qty{36}{\micro\coulomb\per\centi\metre\squared}.
The PUND method was used to distinguish the switching and non-switching contributions to the polarization, and it shows that both samples, as-deposited and annealed, have a switching contribution and that annealing increases this contribution to the polarization by a factor 1.6.
Moreover, the PUND shows that the increase in polarization is also due to the higher non-switching contribution, i.e. capacitive contribution, resulting from a higher relative permittivity.

Using the measurement of dielectric properties as a function of frequency, we show that the dielectric relaxation around \qty{20}{\kHz} at \qty{298}{\K} occurs only when the sample has been annealed.
In addition, the annealed sample exhibits a larger relative permittivity and dielectric losses which is due to a higher ferroelectric behavior of the material and consistent with the PUND measurement.

Impedance spectroscopy as a function of the driving field magnitude, coupled with the Rayleigh analysis, allows us to determine the various contributions to the relative permittivity, lattice, reversible and irreversible domain wall contributions.
For the as-deposited sample, the contribution of the domain wall motions to permittivity is very small, which is due to a low density of the domain walls.
For the annealed sample, the domain wall motion contributions to the permittivity is much higher, which is due to a higher domain wall density.
However, values of the threshold field suggest that the domain wall mobility of the samples remains similar after thermal annealing, therefore no additional defect acting as domain wall pinning centers are induced.

\section*{Data availability}
The data that support the findings of this study are available from the corresponding author upon reasonable request.

\section*{Supplementary Material}
See supplementary material for structural and microstructural analyses of the as-deposited and annealed PMN--35PT thick films.

\section*{Acknowledgments}
This work has been performed with the means of the CERTeM (microelectronics technological research and development center) of French region Centre Val de Loire.
HU and MS acknowledge the Slovenian Research Agency (project J2-3058, bilateral project BI-FR/21-22-PROTEUS-004, young researcher project, research core funding P2-0105) and JSI Director’s fund 2017-ULTRACOOL. 

\section*{Conflict of Interest}
The authors declare no competing financial interest.

\bibliographystyle{aipnum4-2}
\bibliography{biblio_ferro.bib}
\end{document}